\documentclass{article}
\usepackage{graphicx}
\usepackage{url}

\usepackage{breakurl}

\usepackage{arxiv}

\usepackage[utf8]{inputenc} % allow utf-8 input
\usepackage[T1]{fontenc}    % use 8-bit T1 fonts
\usepackage{hyperref}       % hyperlinks
\usepackage{url}            % simple URL typesetting
\usepackage{booktabs}       % professional-quality tables
\usepackage{amsfonts}       % blackboard math symbols
\usepackage{nicefrac}       % compact symbols for 1/2, etc.
\usepackage{microtype}     
\usepackage{subcaption}

\title{Killings of social leaders in the Colombian post-conflict: Data analysis for investigative journalism}

\author{Maria De-Arteaga\thanks{Indicates equal contribution.},\ \  Benedikt Boecking$^{*}$\\ 
  Carnegie Mellon University\\
  Pittsburgh, PA 15213 \\
 }

\begin{document}
\maketitle
%
% The abstract is a short summary of the work to be presented in the article.
\begin{abstract}
%The Colombian post-conflict has been marked by growing concerns regarding the rise killings of social leaders. In this paper we present a data analysis of 
After the peace agreement of 2016 with FARC, the killings of social leaders have emerged as an important post-conflict challenge for Colombia. We present a data analysis based on official records obtained from the Colombian General Attorney's Office spanning the time period from 2012 to 2017. The results of the analysis show a drastic increase in the officially recorded number of killings of democratically elected leaders of community organizations, in particular those belonging to \textit{Juntas de Acci{\'o}n Comunal} [Community Action Boards]. These are important entities that have been part of the Colombian democratic apparatus since 1958, and enable communities to advocate for their needs. We also describe how the data analysis guided a journalistic investigation that was motivated by the Colombian government's denial of the systematic nature of social leaders killings.
%Amongst the challenges that Colombia has faced in the current post-conflict, is the killings of  social leaders. 

%In 2016, Colombia marked the end of an armed conflict that lasted over 50 years. The post-conflict has seen growing concerns regarding the killings of \textit{l{\'i}deres sociales} [social leaders]. However, for a long time the government denied social leaders were likely being targeted due to their roles in social movements. In this paper we analyze officially recorded killings of social leaders,

\end{abstract}



%
% Keywords. The author(s) should pick words that accurately describe the work being
% presented. Separate the keywords with commas.

%
% A "teaser" image appears between the author and affiliation information and the body 
% of the document, and typically spans the page. 
%   \Description{Enjoying the baseball game from the third-base seats. Ichiro Suzuki preparing to bat.}

%
% This command processes the author and affiliation and title information and builds
% the first part of the formatted document.
\maketitle

\section{Introduction}\label{sec:intro}
%the government said the killings were not targeted, what does their data say?

In 2016, Colombia signed a peace agreement with the Fuerzas Armadas Revolucionarias de Colombia (FARC), a Marxist guerrilla known by many for leading the longest insurgency in the world~\cite{leech2011farc}. The year in which the deal was signed grass-root organizations, human rights groups, and journalists started voicing their concerns regarding what was perceived as a growing wave of killings of social leaders~\cite{ElEspectador_homicidios, Semana_panico}. These reports quickly led to fears amongst the Colombian political left, where many saw echoes of the genocide of the political party Union Patriotica~\cite{gomez2007perpetrator,steele2018democracy}, which was founded by FARC during a failed peace process in 1985. The Colombian government's response to those raising alarms about the killings of social leaders was initially--and for a long time--to deny that the victims were being targeted due to their involvement in social movements. High-profile statements include the Defense Vice-Minister declaring that ``\textit{los homicidios de l{\'i}deres sociales no son sistem{\'a}ticos}'' [killings of social leaders are not systematic]~\cite{ElPais_ministro}, and an infamous statement by the Defense Ministry arguing that ``\textit{asesinatos de l{\'i}deres son por l{\'i}os de faldas}'' [killings of (social) leaders are troubles over skirts]~\cite{ElEspectador_ministro}, where ``l{\'i}os de faldas", roughly translated to ``troubles over skirts'', is a dubious popular expression that refers to fights among men over women.

In this context, the authors of this article partnered with local journalists in Cali, Colombia, with the goal of conducting data analysis to guide a journalistic investigation into commonalities and trends in officially recorded killings of social leaders. In this paper, we present the results of the data analysis, discuss the patterns found, and describe how the data analysis guided the journalistic reporting that followed, which resulted in an investigative piece published in October 2018 by El Pais de Cali and CONNECTAS\footnote{\href{https://www.connectas.org/especiales/lideres-en-via-de-extincion/}{https://www.connectas.org/especiales/lideres-en-via-de-extincion/}}.

While ``l{\'ider social}'' [social leader] is a common term in Colombia frequently used by media, government agencies, and the public at large, there is no universally agreed upon definition for it. Therefore, organizations that have devoted themselves to monitoring these events may not always record the same victims. We analyze a dataset of 358 killings of social leaders recorded by the Colombian General Attorney's Office between the years of 2012 and 2017. Social leaders recognized by the government as such include indigenous chiefs~\cite{Semana_ind}, LGBTQ+ activists~\cite{Heraldo}, environmental activists~\cite{vanguardia}, unionists~\cite{Eltiempo_sindical}, among many others. Our findings show that while Colombia has seen a consistent decline in the \textit{total} number of homicides, the number of killings of social leaders as officially recorded by the Colombian General Attorney's Office has been increasing since 2015. We show that the most prominent increase corresponds to the killings of democratically elected leaders of community organizations that represent small territories and are legally recognized entities. 
These organizations include \textit{Juntas de Acci{\'on} Comunal} [community action boards], a type of organization institutionalized in Colombia since 1958~\cite{valencia2010hacia}. The category also includes \textit{Juntas Administradora Local} [local administrative boards] and \textit{Consejos Comunitarios} [community councils], administrative entities of local territories, both of which were created as part of the Colombian Constitution in 1991.

%As such, it is non-trivial to understand whether there is an increase in the number of killings, and if so what may be driving it.  

% the most common of which are \textit{Consejos Comunitarios de Comunidades Negras}, through which black communities administer collective property~\cite{arevalo2001participacion}.%, particularly  prominent in afrocolombian territories.

%Mention the other organizations, find a term that encompasses them all. What do they have in common? They are publicly elected, they represent communities at a local level. These are forms of organization that are stipulated by law, and are an important part of civic democratic organization in Colombia. 

%These are all legally recognized by the government, is a way for territories to organize themselves  and advocate for their needs. Important in the  democratic processes in Colombia. 

%JAC 

%JAL were created in the constitution of 1991

%'CONSEJO COMUNITARIO': A

% 'ASOCIACION COMUNITARIA'
% 'ASOCOMUNAL'

After a review of related work in Section~\ref{sec:rel}, we describe the data used in Section~\ref{sec:data}, followed by a description of data preprocessing and cleaning in Section~\ref{sec:proc}. In Section~\ref{sec:results} we describe the data analysis and present our results, while we use Section~\ref{sec:journ} to discuss how the data analysis guided the journalistic investigation. We close in Section~\ref{sec:conc} with a discussion, conclusions, and future work.

%, including some excerpts of that work

%The growing concerns on the part of both demobilized individuals and Government's initial response "lios de faldas". 

%Colombia signed the peace process, this was shortly followed by an increase in news denouncing the murder of social leaders. 

%Provide details of the news, movements that followed.

\section{Related Work}\label{sec:rel}

The literature on violence in Colombia is vast, and arguably has its origins in the works of Orlando Fals Borda, Eduardo Uma{\~n}a and Germ{\'a}n Guzm{\'a}n--who came to be known as \textit{violentologos}~[violentologists]--a group of Colombian sociologists tasked by the government with studying and documenting the bipartisan war that took place between 1946 and 1966~\cite{guzman1962violencia}. Since then, many more have studied the phenomenon of violence in Colombia~\cite{pecaut1999banality, steele2017democracy}, including the guerrilla movements~\cite{Francisco2008}, the paramilitary violence~\cite{watch2010paramilitaries, ronderos2014guerras}, and the involvement of the government in war crimes~\cite{CARDENAS201364}. Of particular relevance is the work of Centro Nacional de Memoria Hist{\'o}rica, a government institution that has been tasked with documenting the atrocities of war and building historical memory since 2011~\cite{gallon2013desafiando, historica2013basta}.

In 2018, the Human Rights Data Analysis Group published a report on the killings of social leaders in Colombia~\cite{ball2018asesinatos}. The authors use data collected by six organizations in 2016 and 2017 to estimate the total number of killings of social leaders during this period of time, and establish that the total number of victims increased between 2016 and 2017. Importantly~\cite{ball2018asesinatos} find that the estimates of social leader killings do not vary greatly across datasets.  

%\cite{report_HRDAG} A report by HRDAG analyses data of killings in 2016 and 2017 from different non-profits, estimates the number of killings and establishes that there is a high probability that the \textit{true} number of killings increased between 2016 and 2017.

In a working paper,~\cite{prem2018killing} use data from the Colombian nonprofit organization \textit{Somos Defensores} to investigate the killings of social leaders. The authors hypothesise that social leaders were increasingly killed by armed groups excluded from the peace process that wanted to consolidate their power, especially in areas where they took over FARC's illegal activities. In the \textit{Somos Defensores} dataset, \cite{prem2018killing} also find that the category of social leaders targeted the most since the beginning of the ceasefire are local community council leaders. 
%https://blogs.lse.ac.uk/latamcaribbean/2019/01/09/a-violent-peace-killing-social-leaders-for-territorial-control-in-colombia/
\section{Data}
\label{sec:data}
We first describe the official dataset on killings of social leaders in Colombia, obtained from the Colombian General Attorney's Office. We then describe the data we assembled of national homicides spanning the same time period as the social leaders data in order to obtain baseline distributions for different attributes and combinations thereof.

\subsection{Social Leaders Killings Data}
The primary dataset used for this research contains all 358 killings of social leaders recorded by the Colombian General Attorney's Office between 2012 and 2017. It was obtained by our collaborators from the newspaper El Pais de Cali and the journalistic platform CONNECTAS, through a \textit{derecho de petici{\'on}} [right to petition]--a proceeding that enables citizens to request data from the government--. While many nonprofit organizations began collecting data after the peace process, the Colombian General Attorney's Office has been collecting data on killings of social leaders since 2012. This dataset has been made publicly available as part of the journalistic investigation\footnote{\url{https://www.connectas.org/especiales/lideres-en-via-de-extincion/tabla.php}}. 
The original dataset contained 19 variables including the victim's full name, date of death, location (municipality and department), the weapon used, the name of the organization the victim belonged to, and the type of organization among others.%https://www.connectas.org/especiales/lideres-en-via-de-extincion/tabla.php}. %This, combined with the fact that using official data would make it harder for the government to dismiss the findings with arguments concerning data quality, led to us choosing the data by the General Attorney's Office as our main source.  
%'SECCIONAL', 'SEXO', 'NOMBRE VICTIMA', 'APELLIDO', 'NOMBRES','CALIDAD DE LA VICTIMA', 'TIPO DE ORGANIZACION','ORGANIZACION A LA QUE PERTENECE', 'NOMBRE DE LA ORGANIZACION','RANGO DENTRO DE LA ORGANIZACION', 'Date', 'DEPARTAMENTO', 'MUNICIPIO','LUGAR ESPECIFICO', 'ARMA EMPLEADA', 'INDICIADOS O IMPUTADOS','SITUACION JURIDICA IMPUTADOS', 'ORDENES DE PROTECCION', 'FASE PRCESAL'

%, both in terms of the questions that may be asked and the conclusions that may be derived from the results.

%hence we cannot presume that we are getting a complete picture 

%and it is possible that rates of under-reporting vary in different regions of the country.

\subsection{National Homicide Data}
To compare trends in the killings of social leaders with the total number of killings in Colombia, we also use data obtained from the Colombian Police Department, available online through the government's open data website~\footnote{The individual datasets for each year can be obtained at \href{https://www.datos.gov.co}{www.datos.gov.co}. For example, the data for 2017 can be found at \href{https://www.datos.gov.co/Seguridad-y-Defensa/Homicidios-2017/mkw6-468s}{https://www.datos.gov.co/Seguridad-y-Defensa/Homicidios-2017/mkw6-468s} }.  These datasets are only available on a yearly basis and have to be merged into one, which is non-trivial due to inconsistent formatting and changing headers. Amongst the various attributes that the datasets provide, we focus our attention on the date of death, location (municipality and department), and the weapon used. 

\subsection{Data limitations} 
It cannot be assumed that all killings of social leaders are contained within the dataset used, as it is possible that there exists under-reporting. Additionally, the General Attorney's Office does not specify what the data generating process is (who reports each killing, how is it established that someone is a social leader, etc.). %The implication is a simple one: just because a killing does not appear in this  dataset, it does not mean that it did not happen.
Therefore, this dataset does not allow us to paint a complete picture of the trends involving all killings of social leaders, and it does not enable us to establish what trends \textit{do not} exist. Rather, we are finding and reporting patterns that are reflected in the official data. However, as we discuss in Section~\ref{sec:conc}, it is very unlikely that the patterns we discover regarding elected community leaders are reflective of  sample selection bias rather than actual underlying patterns. 

\subsection{Open data does not imply readily usable data}%Making data usable
\label{sec:proc}
Both datasets that were used to drive the investigation required extensive cleaning and preprocessing before allowing for a downstream analysis, despite the national homicide dataset being available via Colombia's open data platform, and the social leaders killings dataset being official data obtained through a right to petition. Misspellings, formatting differences, and idiosyncratic nomenclature were abundant across all attributes. %varying abbreviations or names for the same underlying value
The authors were able to correct most inconsistencies as the underlying values were ultimately identical in their meaning. Only 3 cases in the social leaders dataset had to be discarded from our analysis due to ambiguous date entries\footnote{In the social leaders dataset, 3 entries had to be removed because the correct date could not be determined exactly, although all appeared to have been cases from 2012. We also note that one of these entries that was removed was a JAC leader.}. However, it is important to note that a nontrivial knowledge in programming and natural language processing (NLP) was required to efficiently clean both datasets, a task that the authors believe would have been too time consuming and nearly impossible otherwise. This means that a similarly detailed analysis as the one presented in this paper--especially of trends over time and space--currently cannot be produced by journalists without programming knowledge. This is an important point to consider in the debate of what constitutes ``open data'', and in the discussion around the impact on the lack of equal access to certain skills, e.g. a regional newspaper may have a hard time paying for such services. All of the code the authors created to clean the datasets is available online\footnote{\href{https://github.com/benbo/SocialLeaders}{github.com/benbo/SocialLeaders}} to ensure the cleaned datasets are reproducible from the raw data. %\href{https://github.com/benbo/SocialLeaders}{github.com/benbo/SocialLeaders}}

%After merging of individual files as well as cleaning and aligning column names.
We were able to clean many attributes in both datasets by obtaining all unique values and then mapping them to their respective common value. To simplify this matching, the initial data cleaning step consisted of applying common transforms such as the stripping of accents and removal of excessive white spaces. However, this process became difficult for attributes with high arity such as municipalities and departments. For such attributes we obtained or defined ground truth sets, e.g. the official list of municipalities in Colombia, of which there are $1,122$. We then compared all uncleaned unique values to the ground truth values using a fuzzy matching via string edit distance after which all performed edits could be verified manually, which greatly increased the data cleaning efficiency. 

\section{Data analysis and results}
\label{sec:results}
The analysis is centered around temporal trends, types of leaders killed, and spatio-temporal anomalies. As a first step, we compare national trends of homicides with trends in the recorded killings of social leaders. We then delve into the trends observed for social leaders, trying to better understand the type of organizations that are presumably targeted, and finalize our analysis with spatio-temporal anomaly detection. Code to reproduce all figures and statistics is publicly available~\footnote{\url{https://github.com/benbo/SocialLeaders}}.%\href{https://github.com/benbo/SocialLeaders}{github.com/benbo/SocialLeaders}}.

\subsection{National homicides and killings of social leaders}
\begin{figure}
  %  \centering
    \begin{subfigure}{0.48\textwidth}
    \includegraphics[width=\linewidth]{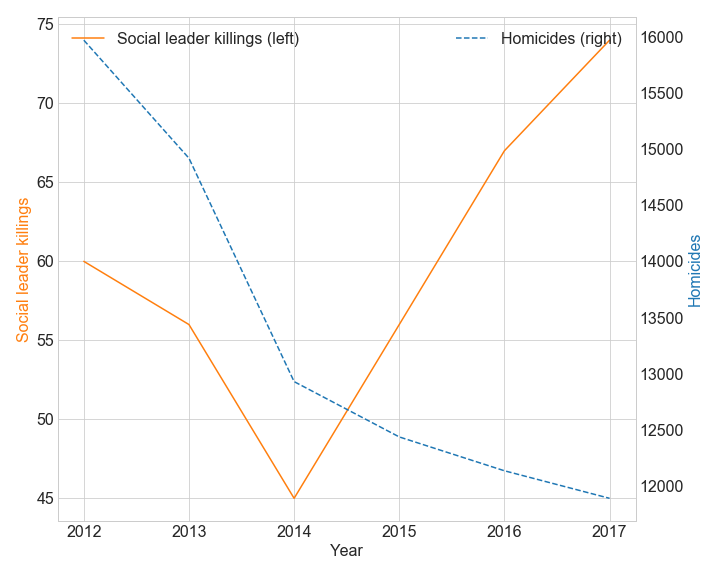}
    \caption{Timeseries of yearly counts of homicides in Colombia \\ and yearly counts of killings of social leaders. }
    \label{fig:time_comp}
    \end{subfigure} 
    \begin{subfigure}{0.48\textwidth}
    \includegraphics[width=\linewidth]{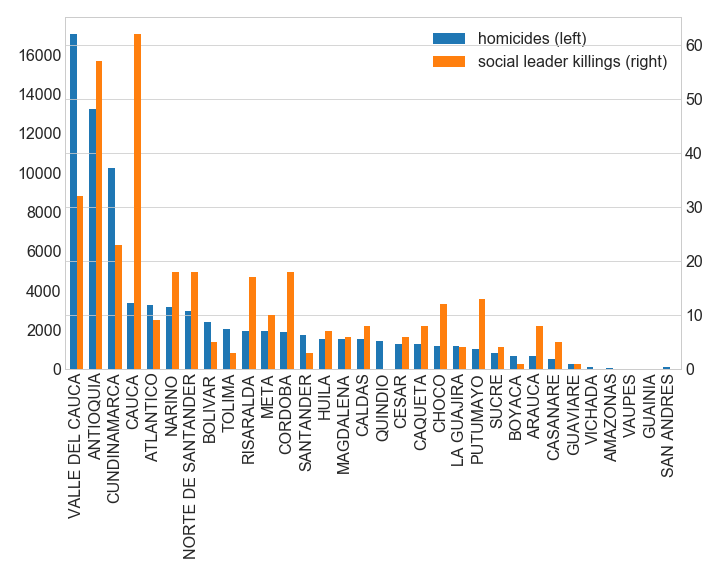}
    \caption{Bar plot showing per state number of homicides and per state number of killings of social leaders side by side. }
    \label{fig:bar_dept}
\end{subfigure}
\caption{Comparison between overall homicides in Colombia and killings of social leaders.}
\end{figure}

A comparison of national homicide data and data on social leader killings offers insights into characteristics and trends in the killings of social leaders.
Figure~\ref{fig:time_comp} shows timeseries for yearly homicide counts in Colombia and yearly numbers of officially recorded social leaders killings from 2012 to 2017.  We observe that while there was a decaying trend in both numbers before 2015, since 2015 the overall number of homicides continued to decrease whereas the number of killings of social leaders started increasing drastically. The difference in the number of killings of social leaders from 2014 to 2017 presents an increase of over 64\%. To put this timeline into perspective, the recent peace process between FARC and the Colombian government began in 2012, FARC declared a unilateral ceasefire in December 2014, and the final peace agreement was reached in August 2016. This was followed by a tumultuous semester during which then President Juan Manuel Santos chose to ratify the deal through a referendum, in which those opposing the agreement won by a thin margin of 50.2\% vs. 49.8\%. A modified agreement was signed--without a referendum--in November the same year.

%offers some evidence of the targeted nature of social leader killings
A comparison of the weapons used shows some clear differences between the two datasets. Amongst the homicides recorded in Colombia between 2012 and 2017, the predominant weapons used with a rate of more than 1\% are 59,643 cases of firearms  (74.29\%), 16,663 knifes/bladed weapons  (20.75\%), and 2,601 cases of blunt force (3.24\%). In the social leaders dataset, the recorded weapons with a rate of more than 1\% are 329 firearms (93.20\%)\footnote{We note that there were 2 instances in the dataset for which a firearm was indicated as the weapon used for the killing, while torture was also indicated in parentheses.}, and 16 knifes/bladed weapons (4.53\%).

The bar plot in Figure~\ref{fig:bar_dept} depicts the number of homicides and the number of killings of social leaders per department. The visualization shows that there are stark differences in the distribution over departments. Of note in particular are the proportional deviations in Cauca and Valle Del Cauca, especially since these are neighboring departments. 

% Homicides data: of which 5725 (8.43\%) were women, 62109 (91.54\%) were men, and 14 (0.02\%) were indicates to have no gender reported. 

%Descriptive stats: Female/Male, Type of weapon (maybe a chi-square test) --disproportionate prevalence of firearms--, although we note we are not controlling for other covariates, demographic distribution.

% JAC emphasis: JAC most prominent. Show this, e.g. a table. #

\subsection{Types of leaders at increasing risk}\label{subsec:types}
\begin{figure}
    \begin{subfigure}[t]{.5\textwidth}\centering
    \includegraphics[width=1.0\columnwidth]{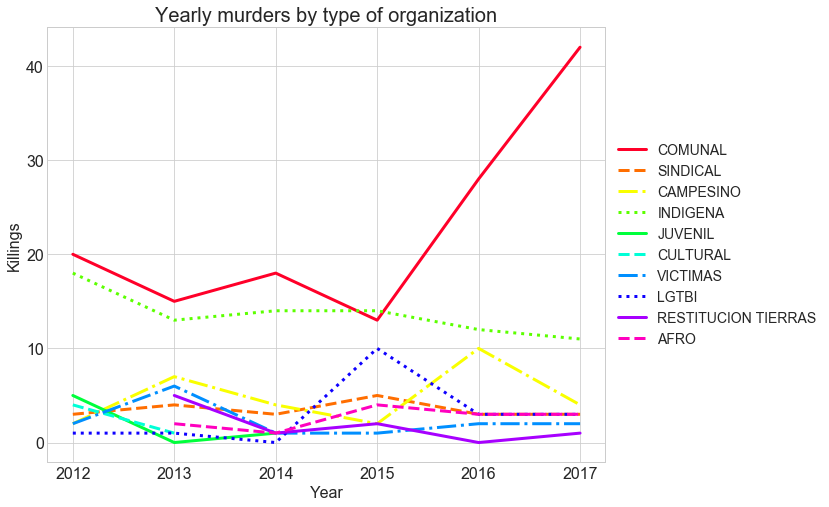}
    \caption{Timeseries of yearly counts of killings of social leaders by type of organization, as typified by the Colombian General Attorney's Office. Yearly ticks on the bottom axis correspond to the end of each year.}
    \label{fig:time_type}
\end{subfigure}
\hspace{1em}
\begin{subfigure}[t]{.5\textwidth}\centering
    \includegraphics[width=0.7\columnwidth]{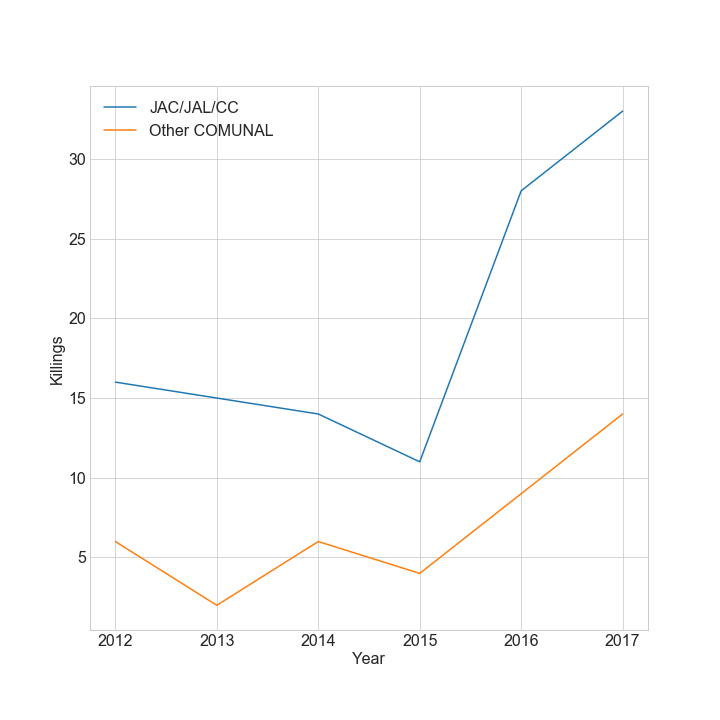}
    \caption{Breakdown of yearly counts of killings of social leaders of the type \textit{comunal}, separated into those belonging to JAC/JAL/CC and social leaders belonging to other types of organization typified as \textit{comunal}. }
    \label{fig:time_JAC}
    \end{subfigure}
    \caption{Timeseries of killings of social leaders by type of organization.}
\end{figure}

 When analyzing the temporal trends of the type of organization to which the victims belonged to, shown in Figure~\ref{fig:time_type}, it becomes clear that the increase in officially recorded killings of social leaders which we observe in Figure~\ref{fig:time_comp} can almost exclusively be attributed to a sharp increase of victims belonging to the category of communal leaders (``comunal''). At first glance, this may not be very telling as it could appear to be a generic category since \textit{l{\'i}der social} and \textit{l{\'i}der comunal} are often used interchangeably in Colombia. %Even if this equivalence is not technically true, our first hypothesis was that this category may have been used as a ``catch-all'', which motivated an analysis of the names of the organizations. 
% \begin{figure}
%     \centering
%     \includegraphics[width=0.65\linewidth]{time_year_tipoorg.png}
%     \caption{Timeseries of yearly counts of killings of social leaders by type of organization, as typified by the Colombian General Attorney's Office. Yearly ticks on the bottom axis correspond to the end of each year.}
%     \label{fig:time_type}
% \end{figure}
However, an analysis of the types of organizations that the social leaders in the ``comunal'' category belong to revealed that a large number are social leaders who hold officially recognized and democratically elected positions. There were 3 designations in particular, namely ``Junta de Acci{\'o}n Comunal'' (JAC), ``Junta Administradora Local'' (JAL), and  ``Consejo Comunitario'' (CC).  We used regular expressions to extract these designations--including abbreviations and variations-- from the fields in the dataset that describe the organization each leader belonged to. The largest group in this category accounting for 104 victims in the cleaned data are leaders of JAC, officially recognized community action boards that represent small territories\footnote{We note that one entry in the dataset that had to be removed due to an ambiguous date was also a JAC leader, bringing the number of JAC leaders in the raw data to 105.}. Eleven victims are leaders of CCs, which are administrative entities for special territories, such as some of the territories where collective property--as opposed to private property--is recognized. Four victims in the ``comunal'' category belonged to JAL, which are administrative entities of small territories present throughout the country. Figure~\ref{fig:time_JAC} shows that these officially recognized and elected leaders make up a large portion of the ``comunal'' category and strongly contribute to the increase in the number of victims since 2015.  

It is also relevant to note that the increase between 2014 and 2015 appears to be of a different nature than the one from 2016 onward. While the overall killings of social leaders as recorded by the General Attorney's office started increasing in 2015, the killings of leaders of the type ``comunal'' only started in 2016, while 2015 saw a peak in killings of LGBTQ+ leaders.
% \begin{figure}
%     \centering
%     \includegraphics[width=0.5\linewidth]{JACJALCC_vs_ComunalOtraSQUARE.png}
%     \caption{Breakdown of yearly counts of killings of social leaders of the type \textit{comunal}, separated into those belonging to JAC/JAL/CC and social leaders belonging to other types of organization typified as \textit{comunal}. }
%     \label{fig:time_JAC}
% \end{figure}

\begin{figure}
    \centering
    \includegraphics[width=0.5\linewidth,trim={10cm 1cm 6cm 1cm},clip]{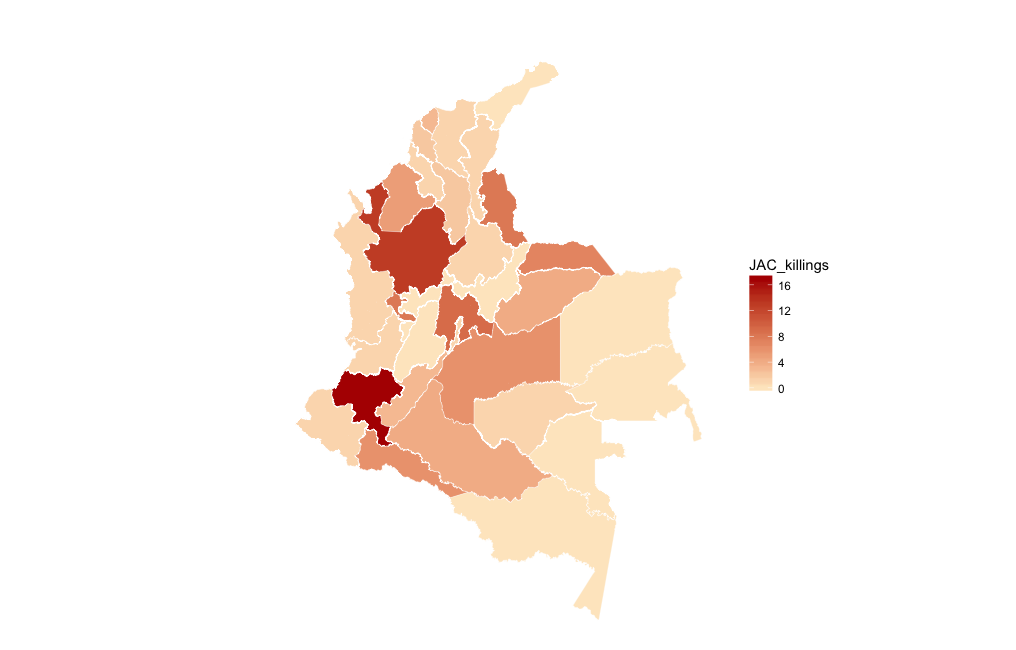}
    \caption{Choropleth map of killings of JAC leaders in Colombia between 2012 and 2017, by department.}
    \label{fig:map}
\end{figure}

The map in Figure~\ref{fig:map} shows a choropleth map of killings of JAC leaders by department. Cauca (17 victims) and Antioquia (13 victims) are the two departments with the highest number of JAC victims\footnote{Additional departments with more than 1 JAC victim: Cundinamarca (9),
Risaralda (8), Norte De Santander (8), Arauca (7), Meta (6), Putumayo (6), Cordoba (5), Caqueta (4), Casanare (4), Huila (4), Atlantico (3), Bolivar (2).}. We further inspect the most recent killings recorded in the data during the last quarter of 2017 to guide the journalistic investigation. It is particularly noticeable that four victims recorded since October 1, 2017 are from Norte de Santander, which corresponds to half of the reported JAC killings in this department and is the highest number of recorded killings during this short time window. In the same time period, two JAC killings each were recorded in each Antioquia and Cauca, and one was recorded in each Caqueta and Nari{\~n}o, amounting to a total of 10 JAC victims during the last quarter of 2017.
\subsection{Spatio-temporal anomalies}
An important part of journalistic reporting is selecting cases to delve into through interviews and chronicles. Several criteria were taken into account in this selection, which are described  in more detail in Section~\ref{sec:journ}. One of these was a spatio-temporal anomaly detection framework~\cite{Dubrawski2011}. The goal here is to find cases in which the counts of a certain type of killing of social leaders deviates from what is expected based on the historical data from a location and what is observed in the rest of the country during a window of time. This enables us to find  locations that experienced a particularly high number of victims in a particular period of time. Another advantage is that the search can be conditioned on attributes of the user's choosing, e.g. to find spatio-temporal anomalies amongst weapons used or gender of the victim. This methodology has previously been applied in order to find patterns of sexual violence in El Salvador~\cite{DeD2016}, to study potential human trafficking activity in connection with public events~\cite{boecking2018quantifying}, and for monitoring of food and agriculture safety~\cite{roure2007study}.  

Scanning for spatio-temporal anomalies, we conditioned on a number of attributes such as gender, weapon used, and a binary JAC indicator. It must be noted that the anomalies the method uncovered were not of the magnitude that has been observed in other domains~\cite{DeD2016,boecking2018quantifying}. This is likely explained by the fact that there is no short window of time for which a given location contains a very high number of killings, and for wider windows of time the observed counts do not constitute a localized pattern and as such are not spatio-temporally anomalous. One noteworthy anomalous pattern --of interest due to our findings of stark increases in JAC victims as discussed in Section \ref{subsec:types}--consisted of three female JAC leaders who were killed in the municipality of Fortul (Arauca) during a window of one month in November 2014. This pattern is particularly interesting given that the majority of recorded killings correspond to men: 13.73\% of the recorded social leaders killed are women, a percentage that increases to 19.23\% when considering only JAC leaders\footnote{Amongst recorded killings of social leaders, 49 (13.73\%) were women, 290 (81.23\%) were men, and 18 (5.04\%) had their gender recorded as ``LGTBI". Upon manually checking the latter, it appears to be the case that transgender victims, as well as any LGBTQ+ leader, had their gender recorded as ``LGTBI'', including gay men, for example. This points to a conflation of gender identity and sexual orientation by those recording the data.}.

%Murder of three female leaders, involved in JAC.

\section{Data-driven reporting}
\label{sec:journ}

%A focus point of the journalistic investigation were the findings of the data analysis, and b
Based on the findings of the data analysis it was determined that the focus of the reporting would be on killings of elected leaders of official local community entities, with a special focus on JAC, as killings of members of this type of organization were particularly prominent. Once the decision had been made, two key considerations remained: where to go and who to interview. %These choices were made in great part driven by the data, but were also influenced by the criteria of the journalist in charge. 

\paragraph{Where to go?}  Based on the trends found in the data, the locations that were chosen for the journalistic investigation are described below. As noted, manual inspection of the cases and the journalists' domain knowledge were also used to inform the choices:
\begin{itemize}
    \item Arauca: Fortul (Arauca) is the municipality where it was found--through use of the spatio-temporal anomaly detection framework--that three female JAC leaders were killed during a time period of less than one month in 2014. This location was selected in order to (1) gain a better understanding of what drove this anomaly, (2) highlight the experience of female leaders, and (3) report on a case from several years ago, to understand how authorities were handling one of these ``cold cases''. 
    \item Cauca: This department was chosen given that it is the one with the highest number of killed JAC leaders. The municipality of El Tambo in Cauca has the highest number of JAC victims (4). Upon manually reviewing cases from Cauca and using their knowledge of the local situation, the lead journalist chose El Tambo as the municipality to visit.
    \item Norte de Santander: This department was chosen given that it had the highest number of recent killings, and this emerging pattern seemed to deviate from the historical trends in this department. Upon reviewing cases in Norte de Santander and using their understanding and domain knowledge of the Colombian post-conflict, the lead journalist chose  Tib\'{u} as the municipality to visit.
\end{itemize}
All travels involved visiting locations with presence of armed groups, and the nature of the reporting was itself a risk factor. Therefore, safety and security guidelines of El Pais de Cali and CONNECTAS were followed--two organizations that have extensive experience in reporting on war and crime. During the security review process of the chosen locations, it was assessed the situation in Norte de Santander around Tib\'{u} presented too much risk. The destination was therefore replaced with Antioquia, the department accounting for the second highest number of JAC victims. Here, the journalist visited the capital--Medellin--which was selected to highlight cases of social leader killings in large cities.    

% location: departamentos with largest number (in the last 2 years) 
% Arauca because of anomaly detection 
% Tibu chosen but too dangerous. 
% Cordoba chosen by journalists.

% potentially a table of states with number of killings and number of most recent killings and number of JAC killings?

\paragraph{Who to interview?} The decision of who to interview was in large part driven by the discovery of the increase in JAC victims in the data recorded by the General Attorney's Office.
\begin{itemize}
    \item Family members and acquaintances of killed leaders: To tell the story of those who were killed, the lead journalist met with family members of killed social leaders of JAC, colleagues who worked with them in the JACs, and other acquaintances.
    \item Threatened JAC leaders: Upon establishing contact with the local JACs in regions where leaders had been killed, it was often discovered by the lead journalist that current presidents and members of JACs had received death threats.  
    \item Researchers employed by nonprofit organizations: There are several nonprofit organizations in Colombia that study the armed conflict, and researchers were interviewed as domain experts. 
    \item Government officials: There are several bodies of government who are responsible for the protection of social leaders. Some of the government officials interviewed include the General Attorney, the National Ombudsman, and the former director of Civic Participation of the Interior Ministry. 
\end{itemize}

%General Attorney

%Defensor Nacional del Pueblo (National Ombudsman)

%Researcher from the NGO \textit{Somos defensores}

%\paragraph{Former Director of Civic Participation of the Interior Ministry}
%What is the role of JAC in politics?

%\paragraph{Social leaders}

\subsection{Journalistic output}

The journalistic investigation was published in October 2018, in print by El Pais de Cali, and online by El Pais de Cali and Connectas\footnote{\url{https://www.connectas.org/especiales/lideres-en-via-de-extincion/}}. The output consists of four chapters and a set of infographics containing the results of the data analysis. The first chapter, titled \textit{Con el asesinato de l{\'i}deres sociales regresa el temor} [With the killings of social leaders, the fear returns], was informed by the interviews with social leaders and their families, who talked about the ways in which the killings bring back the fear that existed during the war and silence the community. The second chapter, \textit{El papel de las JAC} [The role of JAC], discusses the importance of JACs, which are present almost everywhere in Colombia but remain unknown to many Colombians. The third chapter, \textit{El enredo de las cifras} [The mess in the numbers], discusses the issues that stem from the fact that different organizations use different definitions of who is a social leader, and includes interviews with NGOs and government officials on their (often opposing) views about these issues. Finally, the fourth chapter, \textit{Entrevistas} [Interviews], contains three interviews in which two government officials--the General Attorney and the National Ombudsman-- and a researcher of the nonprofit organization Somos Defensores were interviewed regarding the findings of this analysis. 

\section{Discussion and conclusions}
\label{sec:conc}

This paper presents a data analysis of officially recorded killings of social leaders in Colombia between 2012 and 2017. The results show that in 2016 and 2017 there was a concerning increase in the recorded killings of democratically elected leaders of community entities that have a long-standing tradition in Colombia's democracy. Additionally, this paper describes how the data analysis led a journalistic investigation. Identifying the types of leaders who were at a perceived increased risk, and the locations where this trend was particularly concerning, enabled the design of a reporting agenda that centered around the threats faced by leaders of \textit{Junta de Acci{\'on} Comunal} (JAC). The data analysis also informed and provided evidence for interviews with government officials which were  published as part of the journalistic output.    %that was motivated by the government's denial of any systematicity in the killings of social leaders, which at the time translated into a refusal to . 

As discussed in Section~\ref{sec:data}, it is possible that there is under-reporting in the data collected by the General Attorney's Office. Therefore it cannot be discarded that other types of leaders may be at an increased risk of being killed in Colombia. However, it is unlikely that the increase observed in the killings of JAC leaders is just an artifact of such under-reporting and not reflective of an underlying trend. Killings of JAC leaders were recorded since the creation of the data by the General Attorney's Office in 2012. As entities that have been recognized as an official part of the Colombian democratic institutions for more than five decades, it is also unlikely that the killings of their leaders were drastically under-reported until 2016. Additionally, \cite{prem2018killing} also find an increase in killings of what they term ``local community council leaders'' in the data they acquired from the nonprofit organization (NGO) Somos Defensores, and \cite{ball2018asesinatos} find that estimates of social leader killings do not vary greatly across datasets collected by six different organizations. 

The increase in the killings of democratically elected leaders is of great concern in a post-conflict scenario. In particular, JACs are important entities that enable communities to advocate for their needs and have been part of the Colombian democratic apparatus since 1958. The killings of their leaders--especially in rural and marginalized areas--weakens the institutions and hampers citizens' trust in legality. The regions where leaders are at increasing risk, like Cauca, Antioquia, and Norte de Santander, are also the regions where communities have seen some of the most devastating effects of the war. Therefore, democratic institutions in these regions are of particular importance for a lasting peace.

As discussed in Section~\ref{sec:proc}, the obtained data was not usable to compute statistics, time series, maps, and other infographics before it was cleaned and preprocessed, which required computer programming skills. This emphasizes the need to better define what constitutes ``open data'' and to better regulate governments' responsibility in making high quality, usable data available. It also emphasizes some of the risks that stem from unequal access to programming expertise, in which organizations that are fundamental to democracy and development are often unable to pay for important skills and services.

When comparing the trends in national homicides and social leaders' killings from 2012 on, we note that both were decreasing before 2015\footnote{\cite{prem2018killing} also observe a decrease in social leader victims in the same time period in the NGO dataset they use.}. This correlation broke in 2015, when the killings of social leaders started increasing. Analyzing the type of organizations the leaders belonged to, we note that 2015 saw a surge in the recorded killings of LGBTQ+ leaders, while 2016 and 2017 were marked by the sharp increase in the killings of democratically elected leaders--in particular JAC leaders--which we discuss throughout the paper. Developing a better understanding of the increase in the recorded killings of LGBTQ+ leaders in 2015 would be important future work. In particular, we observe that there are zero LGBTQ+  records in 2014, which hints at the possibility that the General Attorney's Office's recognition of who is an LGBTQ+ leader may still be immature, but evolving. 

Future work should study the evolution in the trends of killings of social leaders and the governments' response to this phenomenon. While the new government, which came to power in 2018, does not deny the systematic nature of the phenomenon, the plan it has announced to protect social leaders has drawn heavy criticism from human rights groups~\cite{movice}. In addition to voicing concerns stating that the current plan ignores the guarantees of victims' rights included in the Peace Agreement, it was also noted that the person who was initially chosen as the Director of the new initiative has been accused of systematic killings of civilians to boost military numbers. Of key importance is also the development of a better understanding of the armed groups which are involved in the killings, and the causes driving the violence. Monitoring and understanding targeted violence in the Colombian post-conflict is one of the many things necessary to facilitate peace-building.  

%no indications that would explain this as systematic underreporting that is being corrected.

\section*{Acknowledgement}
We are grateful to Jessica Villamil from El Pais de Cali, Vicente de Arteaga from Early, and to Priscila Hernandez and Carlos Eduardo Huertas from CONNECTAS. Without them, the journalistic investigation would not have been possible.

\bibliographystyle{acm}
\bibliography{bib}

% \appendix
% \label{sec:appendix}

% \section{Gender information}

% In the dataset of national homicides, during this period of time there were 67,848 homicides, of which 5,725 (8.43\%) of the victims were women, 62,109 (91.54\%) were men, and 14 (0.02\%) had no gender reported. The official dataset of social leader killings obtained from the General Attorney's Office recorded 358 killings between 2012 and 2017, of which 49 (13.73\%) were women, 290 (81.23\%) were men, and 18 (5.04\%) had their gender recorded as ``LGTBI". Upon manually checking the latter, it appears to be the case that transgender victims, as well as any LGBTI+ leader, had their gender recorded as ``LGTBI'', including gay victims. This points to a conflation of gender identity and sexual orientation by those recording the data. 

\end{document}